\documentstyle[twocolumn]{jpsj} 
\input epsf.tex
\newcommand{\lsim}
 {\ \raise.35ex\hbox{$<$}\kern-0.75em\lower.5ex\hbox{$\sim$}\ }
\newcommand{\gsim}
 {\ \raise.35ex\hbox{$>$}\kern-0.75em\lower.5ex\hbox{$\sim$}\ }
\def\runtitle{
Absence of Translational Symmetry Breaking 
in Nonmagnetic Insulator Phase 
on Two-Dimensional Lattice with Geometrical Frustration
}
\def\runauthor{Shinji {\sc Watanabe} and Masatoshi {\sc Imada}
}
 
\title
{
Absence of Translational Symmetry Breaking 
in Nonmagnetic Insulator Phase 
on Two-Dimensional Lattice with Geometrical Frustration
}

\author
{
Shinji {\sc Watanabe}\footnote{
E-mail: swata@issp.u-tokyo.ac.jp} and Masatoshi {\sc Imada}
}

\inst
{
Institute for Solid State Physics, 
University of Tokyo, 5-1-5, Kashiwanoha, Kashiwa, Chiba 277-8581
}

\recdate
{
\today
}

\abst
{ 
The ground-state properties of the two-dimensional Hubbard model 
with nearest-neighbor and next-nearest-neighbor hoppings at half filling 
are studied by the path-integral-renormalization-group method. 
The nonmagnetic-insulator phase sandwiched by the 
the paramagnetic-metal phase and the antiferromagnetic-insulator phase 
shows evidence against translational symmetry breaking of 
the dimerized state, plaquette singlet state, staggered flux state, 
and charge ordered state.
These results support that the genuine Mott insulator 
which cannot be adiabatically continued to the band insulator
is realized generically by Umklapp scattering through the 
effects of geometrical frustration and 
quantum fluctuation in the two-dimensional system.
}

\kword
{
nonmagnetic insulator, 
geometrical frustration, 
Mott transition, 
two-dimensional Hubbard model, 
path integral renormalization group method (PIRG)
}

\begin{document}
\sloppy
\maketitle

The Mott transition between the metallic and insulating states 
has been a subject of general interest 
in condensed matter physics~\cite{Mott}.
Many intensive studies have been done 
to clarify the nature of the Mott transition 
by employing analytical and numerical methods~\cite{IFT}.
The present understanding on the Mott transition is that 
almost all the Mott insulators have translational symmetry breakings 
such as AF ordering and dimerization in the ground state 
with only a few exceptions~\cite{IFT}. 
When the translational symmetry is broken, the Brillouin zone is 
folded so that the even number of electrons is contained 
in the unit cell~\cite{Slater}. 
These cases may be adiabatically continued to
the band insulator by the band-gap formation, 
and may not belong to the class of the genuine Mott insulator 
which has the odd number of electrons in the unit cell.

For theoretical models, 
an exception of the Mott insulator 
without Brillouin zone folding 
is found in the ground state of 
the one-dimensional Hubbard model at half filling, 
where neither symmetry breaking nor spin excitation gap 
exists~\cite{LW}. 
The absence of symmetry breaking in the ground state implies that 
a highly quantum-mechanical state is realized. 
Then, the following question arises: 
Does the Mott insulator without translational symmetry breaking 
exist except in one-dimensional systems? 
This has been a long-standing problem 
since Anderson's proposal in 70's~\cite{anderson}.
He originally considered a spin-1/2 model on 
the triangular lattice, 
which may realize the genuine Mott insulator 
by the combined effects of strong magnetic frustration and 
quantum fluctuation, 
although there exists an argument for the presence of 
a magnetic long-ranged order in this system~\cite{trianglr,trianglr2}. 
Recently, the Hubbard model on anisotropic triangular lattice has been studied.
It shows the presence of the nonmagnetic-insulator (NMI) phase near the 
boundary of the metal-insulator transition~\cite{MWI}.

\begin{figure}
\begin{center}
\epsfxsize=7cm \epsfbox{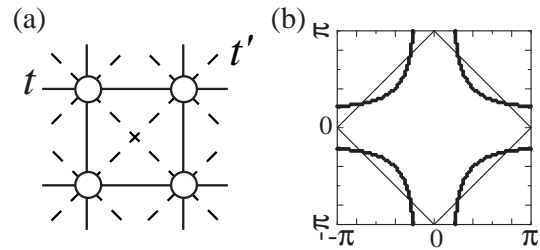}
\end{center}
\caption{
(a)
Lattice structure of the two-dimensional Hubbard model 
on square lattice with transfers $t$ and $t'$.
(b) Fermi surface for $t=1$, $t'=-0.5$ and $U=0.0$ 
at half filling 
on $N=100 \times 100$ lattice (filled circle) and 
magnetic Brillouin zone (solid line).
}
\label{fig:lattice}
\end{figure}

In this letter, we consider the Hubbard model on square lattice (HMSL) 
with nearest and next-nearest neighbor hoppings at half filling 
as a candidate for the genuine Mott insulator. 
Effects of geometrical frustration 
(see Fig.~\ref{fig:lattice}(a)) and quantum fluctuation 
are expected to induce highly quantum-mechanical effects 
in this system. 
Recent weak-coupling renormalization group study suggests 
possible mechanism of insulating gap formation through Umklapp 
scattering without any translational symmetry breaking~\cite{FR}.
Recent path-integral-renormalization-group (PIRG) study 
has clearly shown the presence of an insulating phase without magnetic long-range 
order in intermediate to strong-coupling regime~\cite{KI}. 
It is crucially important to fully understand the nature of the
nonmagnetic Mott insulator phase and also the metallic phase near 
the metal-insulator boundary. 
We study various correlation functions 
to get insight into this problem.
Our results support that the nonmagnetic-insulator phase 
does not show apparent translational symmetry breaking 
near the band-width-control Mott transition.

%
%

The HMSL we consider is 
\begin{eqnarray}
H&=&-\sum_{{\langle i,j \rangle}\sigma}
t_{ij}
\left(
c^{\dagger}_{i\sigma}c_{j\sigma} + c^{\dagger}_{j\sigma}c_{i\sigma}
\right)
+U\sum_{i}n_{i\uparrow}n_{i\downarrow},         
\label{eq:Hamil}
\end{eqnarray}
where $c_{i\sigma}$ $(c^{\dagger}_{i\sigma})$
is the annihilation (creation) operator on the $i$-th site 
with spin $\sigma$ and $n_{i\sigma}=c^{\dagger}_{i\sigma}c_{i\sigma}$
in the $N$-site system. 
The transfer integral is taken as 
$t_{ij}=t$ for the nearest-neighbor sites and 
$t_{ij}=t'$ for the next-nearest-neighbor sites. 
Throughout this letter, we take $t$ as energy unit and 
consider the half-filled case, 
$\sum_{i\sigma}^{N}\langle n_{i\sigma} \rangle /N=1$.

We have applied the PIRG method~\cite{PIRG1,PIRG2} to eq.~(\ref{eq:Hamil}).
This algorithm starts from and improves the mean-field Hartree-Fock solution 
by increasing the dimension of truncated Hilbert space 
in non-orthogonal basis optimized by the path-integral operation. 
A variance extrapolation of improved wave function has been taken 
to reach the true ground state of finite systems in a controlled way.
This method does not have difficulties as in previous methods such as 
the negative sign problem in the quantum Monte Carlo calculations.
The PIRG calculations have been done in the systems up to 
$N=12 \times 12$ lattices under the periodic boundary condition.

\begin{figure}
\begin{center}
\epsfxsize=8cm \epsfbox{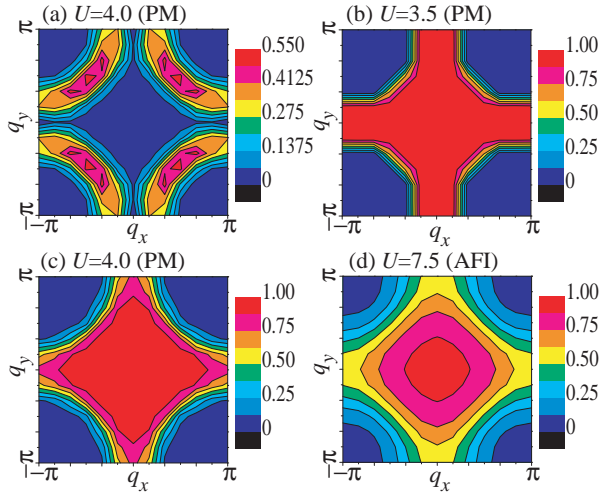}
\end{center}
\caption{
Contour plots of 
$|\nabla_{\bf q} n({\bf q})|$ for (a)$U=4.0$ and 
$n({\bf q})$
for 
(b)$U=3.5$, (c)4.0, and (d)7.5 
for $t=1$, $t'=-0.5$ 
on the $N=12\times 12$ lattice at half filling.
PM and AFI represent paramagnetic metal and antiferromagnetic insulator, 
respectively. 
}
\label{fig:FermiD}
\end{figure}

First we discuss the change of the Fermi surface (FS). 
The FS can be defined by the $\bf q$ points where 
the momentum distribution 
$n({\bf q})=\frac{1}{2N}\sum_{\sigma}\sum_{i,j} 
\langle c_{i\sigma}^{\dagger}c_{j\sigma}\rangle
{\rm e}^{{\rm i}{\bf q}\cdot({\bf R}_{i}-{\bf R}_{j})}
$
exhibits a sharp discontinuity. 
We show 
the contour plots of $|\nabla_{\bf q} n({\bf q})|$ in Fig.~\ref{fig:FermiD}(a) 
to estimate the FS  
for $t=1$ and $t'=-0.5$ on the $N=12 \times 12$ lattice. 
The large values of $|\nabla_{\bf q} n({\bf q})|$ 
specify the possible location of the FS in the momentum space. 
We see that $|\nabla_{\bf q} n({\bf q})|$ has larger values 
around ${\bf q}=(\pi/2,\pi/2)$ 
than those around ${\bf q}=(0,\pi)$ and $(\pi,0)$. 
This indicates that renormalization of electrons 
occurs around both ends of the arc of the FS  
more prominently rather than around the center of the arc. 
Although it is difficult to demonstrate the validity of the 
Luttinger's theorem within the present analysis 
due to the finite-size effects, 
we note that there exists the following tendency concerning the shape  
of the FS: 
As $U$ increases, the FS around $q=(\pi/2,\pi/2)$ 
tends to expand to the $q=(\pi,\pi)$ point 
and the FS around the ends of the arc tends to 
shrink to the ${\bf q}=(0,\pi)$ and $(\pi,0)$ points. 
This can be also seen from the contour plots of $n({\bf q})$ 
in Fig.~\ref{fig:FermiD}(b) $U=3.5$, (c)$U=4.0$ and (d)$U=7.5$. 
The large values of $|\nabla_{\bf q} n({\bf q})|$ 
roughly correspond to the yellow area of the contour plot of $n({\bf q})$
as seen from Fig.~\ref{fig:FermiD}(a) and Fig.~\ref{fig:FermiD}(c). 
The metal-insulator transition is estimated at 
$U_{\rm c1}=4.75\pm0.25$
by the calculations of charge gap and double occupancy, and 
the antiferromagnetic transition specified by 
vector ${\bf Q}=(\pi,\pi)$ 
in spin correlation function, 
$
S({\bf q})=
\frac{1}{3N}
\sum_{i,j}^{N}
\langle {\bf S}_{i}\cdot {\bf S}_{j}\rangle
{\rm e}^{{\rm i}{\bf q}\cdot{({\bf R}_{i}-{\bf R}_{j})}}, 
$
is estimated at $U_{\rm c2}=7.25\pm0.25$~\cite{KI}. 
Although the FS does not exist in the insulator phase, 
we also show here the contour plot of $n({\bf q})$ in the 
antiferromagnetic-insulator (AFI) phase
in Fig.~\ref{fig:FermiD}(d) 
to see the tendency of the deformation of the contour line of the Fermi level.
From Figs.~\ref{fig:FermiD}(b)-(d), 
we see that the FS deforms from the $U=0$ FS 
toward the magnetic Brillouin zone specified by 
the vector ${\bf Q}$, which are 
represented as the filled circles and the solid line, respectively 
in Fig.~\ref{fig:lattice}(b).
Namely, the FS deforms toward the perfect nesting 
as $U$ increases. 
This tendency has been predicted by 
the renormalization-group study on the HMSL 
in the weak-coupling regime~\cite{OMKM}. 

%
%

\begin{figure}
\begin{center}
\epsfxsize=5.3cm \epsfbox{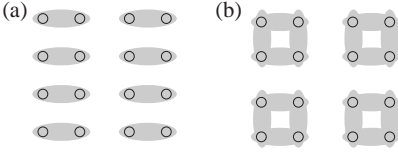}
\end{center}
\caption{
Schematic pictures of 
(a) the columnar-dimerized state and 
(b) the plaquette-singlet state. 
}
\label{fig:dimpla}
\end{figure}

Next we discuss the nature of the NMI phase. 
In the NMI phase for $U_{\rm c1}<U<U_{\rm c2}$, $S({\bf q})$ has 
broad peaks at incommensurate positions 
shifted from ${\bf q}=(\pi,\pi)$ toward 
${\bf q}=(0,{\pm}\pi)$ and $({\pm}\pi,0)$ directions, 
in contrast to the AFI phase for $U>U_{\rm c2}$ 
with the commensurate peak at ${\bf q}={\bf Q}$~\cite{KI}. 
A remaining question to be clarified 
is whether some other translational symmetry is 
broken, or not in the NMI phase. 
To examine the nature of the NMI phase 
let us consider the $U/t \to \infty$ limit of the phase diagram. 
In the strong-correlation limit, the HMSL becomes the $J_{1}$-$J_{2}$ 
Heisenberg model in the leading order: 
$
\tilde{H}
=
 J_{1}\sum_{\langle i,j \rangle}{\bf S}_{i}\cdot{\bf S}_{j}
+J_{2}\sum_{\{ i,j \}}{\bf S}_{i}\cdot{\bf S}_{j},
$
where 
$\langle \rangle$ and $\{ \}$ represent the nearest-neighbor
sites and next-nearest-neighbor sites, respectively. 
Here, $J_{1}=4t^{2}/U$ and $J_{2}=4t'^{2}/U$ are both antiferromagnetic
interactions so that magnetic frustration arises. 
For $J_{2}/J_{1}\lsim 0.4$ the Neel order with peak structure in $S({\bf q})$ 
at ${\bf q}=(\pi,\pi)$ was proposed 
while for $J_{2}/J_{1}\gsim 0.6$ the collinear order with 
${\bf q}=(0,\pi)$ or $(\pi,0)$ peak in $S({\bf q})$ was proposed. 
However, as for the intermediate region of $0.4\lsim J_{2}/J_{1}\lsim 0.6$, 
no definite conclusion has been drawn on the nature of the ground state: 
The possibility of the columnar-dimerized state~\cite{dimer,sushkv} 
(see Fig.~\ref{fig:dimpla}(a)), the 
plaquette singlet state~\cite{plasglt} (see Fig.~\ref{fig:dimpla}(b)) 
and the resonating-valence-bond~\cite{RVB} state was discussed. 
The AFI and the NMI phases found in the HMSL can be 
connected to the Neel-ordered and the NM phases of 
the $J_{1}$-$J_{2}$ Heisenberg model, respectively, 
in the limit $U/t \to \infty$. 
Therefore, the NMI phase in the phase diagram of the HMSL 
could have some connection to the region $0.4 \lsim J_{2}/J_{1} \lsim 0.6$.

%
%

We first note that it is in principle hard to 
prove the existence of the genuine Mott insulator 
because it is difficult to prove the absence 
of all the possible translational symmetry breakings. 
To get insight into this problem, we examine 
every symmetry breaking ever proposed, which is likely to occur 
in the NMI phase.



\begin{figure}
\begin{center}
\epsfxsize=8cm \epsfbox{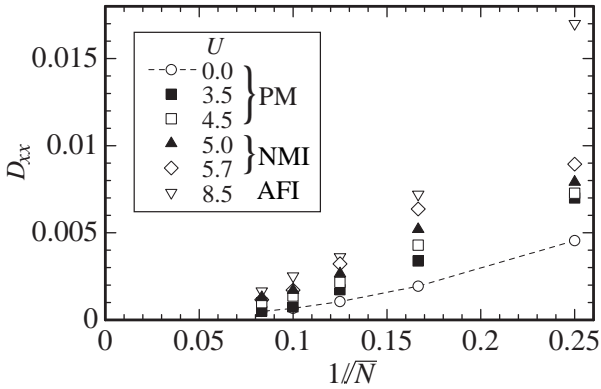}
\end{center}
\caption{
The system-size dependence of 
the dimer-correlation functions 
for $t=1.0$, $t'=-0.5$ and $U=0.0$ (open circle), 
3.5 (filled square), 4.5 (open square), 5.0 (filled triangle), 
5.7 (open diamond) and 8.5 (open triangle) 
at half filling on $N=4\times4, 6\times6, 8\times8, 
10\times10$ and $12\times12$ lattices. 
}
\label{fig:dimer}
\end{figure}

To examine the possibility of the dimerized state 
and the plaquette state, 
we have calculated the dimer-correlation function 
$D_{\alpha\beta}$ for $\alpha,\beta=x,y$ defined by 
$
D_{\alpha\beta}= 
\left\langle O_{\alpha} O_{\beta} \right\rangle, 
$
%
where 
$
O_{\alpha}=
\frac{1}{N}\sum_{i=1}^{N} 
(-1)^{i}{\bf S}_{i}\cdot{\bf S}_{i+{\hat{\alpha}}}
$,
and $\hat{\alpha}$ represents the unit vector 
in the $\alpha$(=$x$ or $y$) direction. 
Figure~\ref{fig:dimer}
shows the system-size dependence of $D_{xx}$ 
for $t'=-0.5$, and 
$U=0.0$, $3.5$, $4.5$, $5.0$, $5.7$ and $8.0$ 
on the $N=4\times4$, $6\times6$, $8\times8$, $10\times10$ 
and $12\times12$ lattices.
We note that the dimer correlation is enhanced in comparison with 
$U=0$ results and it indicates that short-ranged correlation develops. 
The extrapolated values of $D_{xx}$ 
to the thermodynamic limit, however, seem to vanish for 
all interaction strength calculated here. 
These results indicate 
that the columnar-dimerized state and the plaquette-singlet state 
(see Fig.~\ref{fig:dimpla}) proposed as a candidate for the NMI phase 
with broken translational symmetry are not realized in the present system.
The $D_{xy}$ 
correlations have about 2- or 3-orders of magnitude 
smaller values than the $D_{xx}$ correlations for 
each system size and the extrapolation to the thermodynamic limit 
also shows no realization of the $D_{xy}$-type ordering.

%
%


Next, we examine symmetry breaking of various density waves 
proposed in the literature~\cite{Affleck}.
We calculate 
the current-correlation function defined by
%
$$
C_{\alpha}({\bf q})=
\left\langle J_{\alpha}({\bf q})J_{\alpha}^{\dagger}({\bf q}) 
\right\rangle, 
$$ 
where 
$
J_{\alpha}({\bf q})=\frac{1}{N}\sum_{{\bf k},\sigma}
c^{\dagger}_{{\bf k},\sigma}c_{{\bf k}+{\bf q},\sigma}f_{\alpha}({\bf k})
$,
with 
$f_{1}({\bf k})=\cos(k_{x})+\cos(k_{y})$, 
$f_{2}({\bf k})=\cos(k_{x})-\cos(k_{y})$, 
$f_{3}({\bf k})=2\cos(k_{x})\cos(k_{y})$, 
and 
$f_{4}({\bf k})=2\sin(k_{x})\sin(k_{y})$. 

\begin{figure}
\begin{center}
\epsfxsize=7.6cm \epsfbox{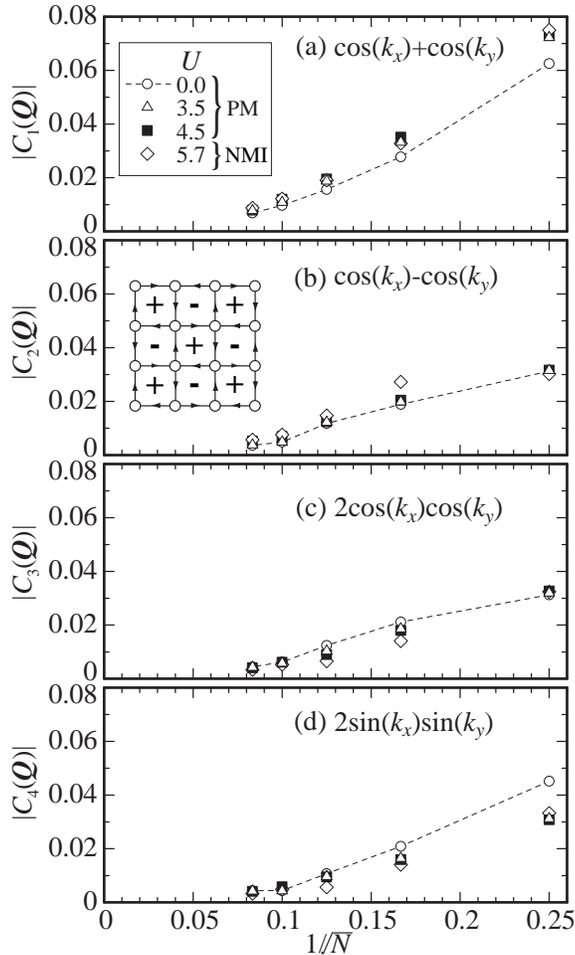}
\end{center}
\caption{
The system-size dependence of 
current correlations $|C_{\alpha}({\bf Q})|$ 
at ${\bf Q}=(\pi,\pi)$
with (a) $f_{1}=\cos(k_{x})+\cos(k_{y})$ 
(b) $f_{2}=\cos(k_{x})-\cos(k_{y})$,
(c) $f_{3}=2\cos(k_{x})\cos(k_{y})$, 
and 
(d) $f_{4}=2\sin(k_{x})\sin(k_{y})$
for $t=1.0$, $t'=-0.5$ and $U=0.0$ (open circle), $3.5$ (open triangle), 
$4.5$ (filled square) and $5.7$ (open diamond)
at half filling on 
$N=4 \times 4$, $6 \times 6$, $8 \times 8$, $10 \times 10$, 
and $12 \times 12$ lattices. 
Inset of (b) illustrates schematic picture of the staggered-flux state.
}
\label{fig:current}
\end{figure}

Among finite ${\bf q}$, we have confirmed that 
the peak structure in $C_{\alpha}({\bf q})$ for $\alpha=1,2,3$ and 4 
appears at ${\bf q}=(\pi,\pi)={\bf Q}$ 
for $t'=-0.5$, and $U=0.0$, $3.5$, $4.5$, and $5.7$. 
Here $C_{2}({\bf Q})$ corresponds to 
the so-called staggered-flux state~\cite{Affleck}
(see inset of Fig.~\ref{fig:current}(b)).
Figure~\ref{fig:current} shows 
the system-size dependence of $|C_{\alpha}({\bf Q})|$, for $t'=-0.5$, and 
$U=0.0$, $3.5$, $4.5$ and $5.7$ 
on the $N=4\times4$, $6\times6$, $8\times8$, $10\times10$ 
and $12\times12$ lattices.
If the long-ranged order of these types occurs, 
$|C_{\alpha}({\bf Q})|$ should have 
a finite value in the thermodynamic limit. 
Although 
$|C_{\alpha}({\bf Q})|$ for $\alpha=1$ and $2$ are enhanced in comparison 
with $U=0$ results, all current correlations calculated here 
seem to vanish for $N \to \infty$ 
and the correlations remain short ranged. 
These results suggest that the symmetry breaking 
specified by
the order parameter, $J_{\alpha}({\bf q})$ for $\alpha=1,..,4$
does not occur in the NMI phase.

%
%

\begin{figure}
\begin{center}
\epsfxsize=8cm \epsfbox{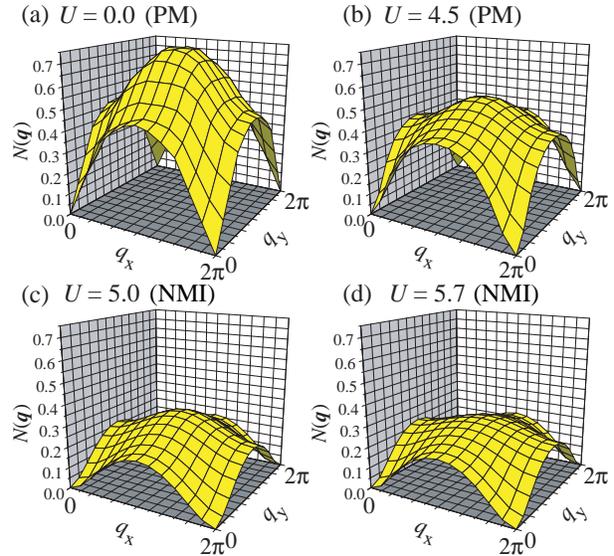}
\end{center}
\caption{
The charge correlation functions for $t=1.0$, $t'=-0.5$, and 
(a)$U=0.0$, (b)$4.5$, (c)$5.0$ and (d)$5.7$ at half filling 
on the $N=12 \times 12$ lattice.
}
\label{fig:charge}
\end{figure}

\begin{figure}
\begin{center}
\epsfxsize=8cm \epsfbox{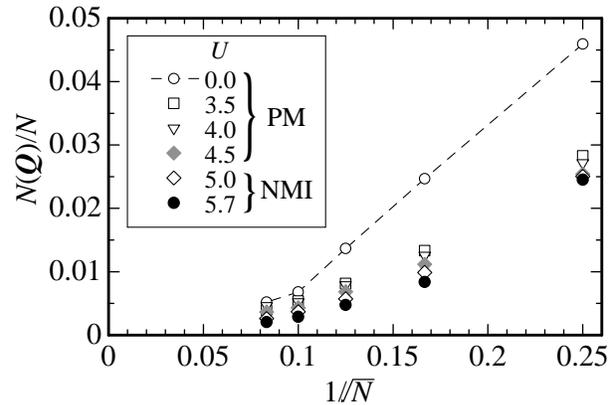}
\end{center}
\caption{
The system-size dependence of 
the peak value of charge correlation functions 
for $t=1.0$, $t'=-0.5$ and $U=0.0$ (open circle), 
3.5 (open square), 4.0 (open triangle), 4.5 (filled diamond), 
5.0 (open diamond) and 5.7 (filled circle) 
at half filling on $N=4\times4, 6\times6, 8\times8, 
10\times10$ and $12\times12$ lattices. 
}
\label{fig:peakNq}
\end{figure}

We have also calculated the charge-correlation function 
defined by 
%
\begin{eqnarray}
N({\bf q})=\frac{1}{N}\sum_{i,j}^{N}
\left(
\langle n_{i}n_{j} \rangle
-
\langle n_{i} \rangle
\langle n_{j} \rangle
\right)
{\rm e}^{{\rm i}{\bf q}\cdot({\bf R}_{i}-{\bf R}_{j})},
\nonumber
\end{eqnarray}
with $n_{i}=\sum_{\sigma}c_{i\sigma}^{\dagger}c_{i\sigma}$. 
We show $N({\bf q})$ in Fig.~\ref{fig:charge} 
for (a)$U=0.0$, (b)4.5, (c)5.0, and (d)5.7 for $t'=-0.5$ 
on the $N=12\times 12$ lattice.
No conspicuous peak structure develops even in the 
NMI phase such as $U=5.0$ and 5.7. 
Figure~\ref{fig:peakNq} shows 
the system-size dependence of $N({\bf Q})/N$, for $t'=-0.5$, and 
$U=0.0$, $3.5$, $4.5$, $5.0$ and $5.7$ 
on the $N=4\times4$, $6\times6$, $8\times8$, $10\times10$ 
and $12\times12$ lattices.
The extrapolation of maximum values of $N({\bf q})$ 
to $N \to \infty$  goes to zero for all interaction strength 
calculated here. 
The absence of prominent peak structure in $N({\bf q})$ 
in various system sizes 
excludes the possibility of incommensurate charge orderings.
Thus, we conclude that the charge order does not occur in the NMI phase.

%
%

%
%

In summary, we have studied the ground-state properties of the 
Hubbard model on the square lattice 
with nearest-neighbor and next-nearest-neighbor 
hoppings at half filling by using the PIRG method. 
We show that the renomalization of electrons occurs 
in the metallic phase drastically around 
${\bf q}=(0,\pi)$ and $(\pi,0)$ rather than around ${\bf q}=(\pi/2,\pi/2)$. 
There exists a tendency that 
the Fermi surface deforms toward the magnetic Brillouin zone 
as $U$ increases, namely, toward the perfect nesting. 
In the nonmagnetic-insulator phase sandwiched by the paramagnetic-metal 
and the antiferromagnetic-insulator phases, 
the calculations of dimer-, current-, and charge-correlation functions 
show no evidence of symmetry breakings such as the 
columnar-dimerized state, plaquette-singlet state, staggered-flux state 
(d-density-wave state), 
s-density-wave state, and charge-ordered state. 
Although a possibility 
of some other symmetry breakings 
cannot be excluded, 
our results support that the Mott insulator 
without translational symmetry breaking 
is realized in the nonmagnetic-insulator phase, 
where the translational symmetry breakings undergo quantum melting.
This implies that the genuine Mott insulator
which can not be adiabatically continued to the band insulator 
is realized by Umklapp scattering via the effects of 
geometrical frustration and quantum fluctuation 
near the metal-insulator phase boundary.

%
We thank S. Sachdev for discussions.
The work is supported by the `Research for the Future' program from the 
Japan Society for the Promotion of Science under grant number 
JSPS-RFTF97P01103.
A part of our computation has been done at the supercomputer center 
in the Institute for Solid State Physics, University of Tokyo.



\begin{thebibliography}{99}
\bibitem{Mott} N. F. Mott and R. Peierls: Proc. Phys. Soc. London. 
A{\bf 49} (1937) 72. 
\bibitem{IFT} For review, see M. Imada, A. Fujimori and Y. Tokura: 
Rev. Mod. Phys. {\bf 70} (1998) 1039.
\bibitem{Slater} J. C. Slater: Phys. Rev. {\bf 82} (1951) 538.
\bibitem{LW} E. H. Lieb and F. Y. Wu: Phys. Rev. Lett. {\bf 20}
(1968) 1445. 
\bibitem{anderson} P. Fazekas and P. W. Anderson: Philos. Mag. 
{\bf 30} (1974) 423. 
\bibitem{trianglr} B. Bernu, {\it et al}.: Phys. Rev. B {\bf 50} 
(1994) 10048.
\bibitem{trianglr2}
L. Capriotti, A. E. Trumper and S. Sorella: Phys. Rev. Lett. {\bf 82} 
(1999) 3899. 
\bibitem{MWI} H. Morita, S. Watanabe and M. Imada: 
preprint(cond-mat/0203020).
\bibitem{FR} 
N. Furukawa and T. M. Rice: Phys. Rev. Lett. {\bf 81} 
(1998) 3195; 
C. Honerkamp, {\it et al}. :Phys. Rev. B {\bf 63} 
(2001) 35109.
\bibitem{KI}  T. Kashima and M. Imada: J. Phys. Soc. Jpn. {\bf 70} 
(2001)  3052.
%
%
\bibitem{PIRG1} M. Imada and T. Kashima: J. Phys. Soc. Jpn. {\bf 69} 
(2000)  2723.
\bibitem{PIRG2} T. Kashima and M. Imada: J. Phys. Soc. Jpn. {\bf 70} 
(2001)  2287.
\bibitem{OMKM} T. Ogawa, {\it et al}. : Physica B {\bf 312-313} 
(2002) 525. 
\bibitem{dimer} N. Read and S. Sachdev: Phys. Rev. Lett. {\bf 66} 
(1991) 1773. 
\bibitem{sushkv} O. P. Sushkov, J. Oitmaa and Z. Weihong: 
Phys. Rev. B {\bf 63} (2001) 104420.
\bibitem{plasglt} M. E. Zhitomirsky and K. Ueda: 
Phys. Rev. B {\bf 54} (1996) 9007. 
\bibitem{RVB} L. Capriotti, {\it et al}. : Phys. Rev. Lett. {\bf 27} 
(2001) 97201. 
\bibitem{Affleck} I. Affleck and J. B. Marston: Phys. Rev. B {\bf 37}
(1988) 3774. 
\end{thebibliography}
\end{document}